\documentclass[a4paper,english]{article}

\usepackage{fullpage}
\usepackage{amsmath,amssymb,amsthm,amstext}
\usepackage{authblk}
\usepackage[inline]{enumitem}
\usepackage{graphicx}
\usepackage{wrapfig}
\usepackage{subcaption}
\usepackage{scrextend}

\newcommand{\enquote}[1]{\emph{``#1''}}

\renewcommand{\vec}[1]{\overrightarrow{#1}}

\usepackage{hyperref}
\hypersetup{colorlinks,breaklinks,
            urlcolor=[rgb]{0,.4,.8},
            linkcolor=[rgb]{0.3,0.3,.8},
            citecolor=[rgb]{.8,.2,.2}}

\graphicspath{{figures/}}

\begin{document}
\title{%
Experimental Analysis of the Accessibility of Drawings with Few Segments\thanks{%
This work was partially funded by the German Research Foundation (grant SCHU 2458/4-1).
W. Meulemans is funded by the Netherlands eScience Center (NLeSC, grant 027.015.G02).}}

\author[1]{Philipp Kindermann}
\author[2]{Wouter Meulemans}
\author[3]{Andr\'e Schulz}

\affil[1]{David R. Cheriton School of Computer Science, University of Waterloo, \texttt{philipp.kindermann@uwaterloo.ca}}
\affil[2]{Dept. of Mathematics and Computer Science, TU Eindhoven, \texttt{w.meulemans@tue.nl}}
\affil[3]{LG Theoretische Informatik, FernUniversit\"at in Hagen, \texttt{andre.schulz@fernuni-hagen.de}}

\maketitle

\begin{abstract}
The visual complexity of a graph drawing is defined as
the number of geometric objects needed to represent all its edges.
In particular, one object may represent multiple edges, e.g., one needs
only one line segment to draw two collinear incident edges.
We investigate whether drawings with few segments have a better aesthetic appeal and help the user
to assess the underlying graph.
We design a user study that investigates two different graph types
(trees and sparse graphs), three different layout algorithms for trees, and
two different layout algorithms for sparse graphs. We asked the participants to
give an aesthetic ranking on the layouts and to perform a furthest-pair or shortest-path task on the
drawings.
\end{abstract}


\section{Introduction}
\label{sec:introduction}

Algorithms for drawing graphs try to optimize (or give a guarantee on) certain formal quality measures. Typical
measures include area, grid size, angular resolution, number of crossings, and number of bends.
While each of these criteria is well motivated, we have no guarantee that we get a clear and legible drawing by
optimizing only one of the measures.
This is caused by most measures competing with each other, implying the best score according to one metric may require sacrificing another.
For
example, it is known that certain planar graphs cannot be drawn with good angular resolution and polynomial area~\cite{MP94}.
The question arises how we can select an appropriate algorithm for a given graph drawing task. Instead of
relying on a combinatorial or geometric measure of the drawing, one could also evaluate the
results of the algorithms by measuring the efficiency of tasks carried out by the observer. Another
option would be to just ask observers which drawing they considers \enquote{nicer}. By conducting such
experiments we also hope to learn something about the formal measures.
The goal is to identify
formal measures and algorithms that are particularly suitable for typical tasks performed by using a graph visualization.

A path consisting of several edges may be drawn as a single segment if the edges happen to align their direction.
Although the path may contain many edges, this can be counted as only one segment in the drawing.
The total number of such segments is known as the \emph{visual complexity}
of the drawing. Instead of straight-line segments, one could also use other geometric objects to draw
paths. One option that has been introduced by Schulz~\cite{s-dgfa-JGAA15} is using circular arcs.
In this paper our focus lies with drawings using segments.

It is an open question whether a small number of segments is a good quality measure for graph drawings.
We present a study that investigates how drawings with few segments are perceived by the observer in contrast to other drawing styles.
In other words, we want to find out if this design criterion makes drawings more aesthetically appealing for the
observer and/or if they are helpful for executing tasks. The main difficulty is that we cannot control
the visual complexity of a drawing while keeping other quality measures fixed.
One way to avoid this problem is to adapt
existing algorithms in such a way that we can reduce the number of segments in the final drawing without
changing the layout \enquote{style} of the existing drawing too much.

In our study, we focus on two graph classes. The first class is that of (rooted) trees\footnote{In the remainder, we use ``trees'' to refer to rooted trees.}, for which many drawing algorithms
are known. It is not hard to see that every tree can be drawn with $n_\text{odd}/2$ segments,
where $n_\text{odd}$ denotes the number of odd-degree nodes in the tree~\cite{dujmovic2007}. It is unknown however, if every tree
can be drawn with $n_\text{odd}/2$ segments using only a polynomial grid size. We use a heuristic based on
the algorithm of H\"ultenschmidt et al.~\cite{hkms-dttfg-wg17} that draws a tree with minimal
visual complexity and quasi-polynomial area and compare its drawing in the user study against drawings
of other algorithms. In particular, we use the algorithms of Walker II~\cite{w-npagt-SPE90} and
of Rusu et al.~\cite{ryc-psgda-IV08} as alternatives.
The former mimics the standard style in which trees are
typically drawn in the computer science literature. The latter aims to draw trees with good
angular resolution on a small grid.

The second class of graphs we consider consists of sparse but not necessarily planar graphs, as provided by the \emph{ROME
library}~\cite{wdglttv-ecfgd-CG97}. In this setting, it is even harder to control more than one formal measure.
We therefore selected only two algorithms to compare. The first is the popular Fruchterman-Reingold spring
embedder algorithm~\cite{fr-gdfdp-SPE91}. The second is an adaptation of it, aimed at reducing the visual complexity as measured by the number of segments: this is done by adding constraints that force certain edges to be collinear and hence form a straight-line segment. As argued also above, we introduce this new adaptation in order
to generate drawings that have a \enquote{similar feel} but use fewer segments.

We selected two tasks for the participants to evaluate the drawings presented to them. The first task
addresses the question which of the drawings are aesthetically more appealing to the participant. In the second
task, we asked the participants to answer questions. For the trees, we asked to identify pair of nodes realizing the
largest distance; for the sparse graphs we asked to select a shortest path between two designated vertices.
The user study was implemented as a voluntary online questionnaire in order to reach a significant number of participants.

\section{Algorithms}

\begin{figure}[t]
  \begin{subfigure}[b]{.4\textwidth}
    \centering
    \includegraphics[scale=.2]{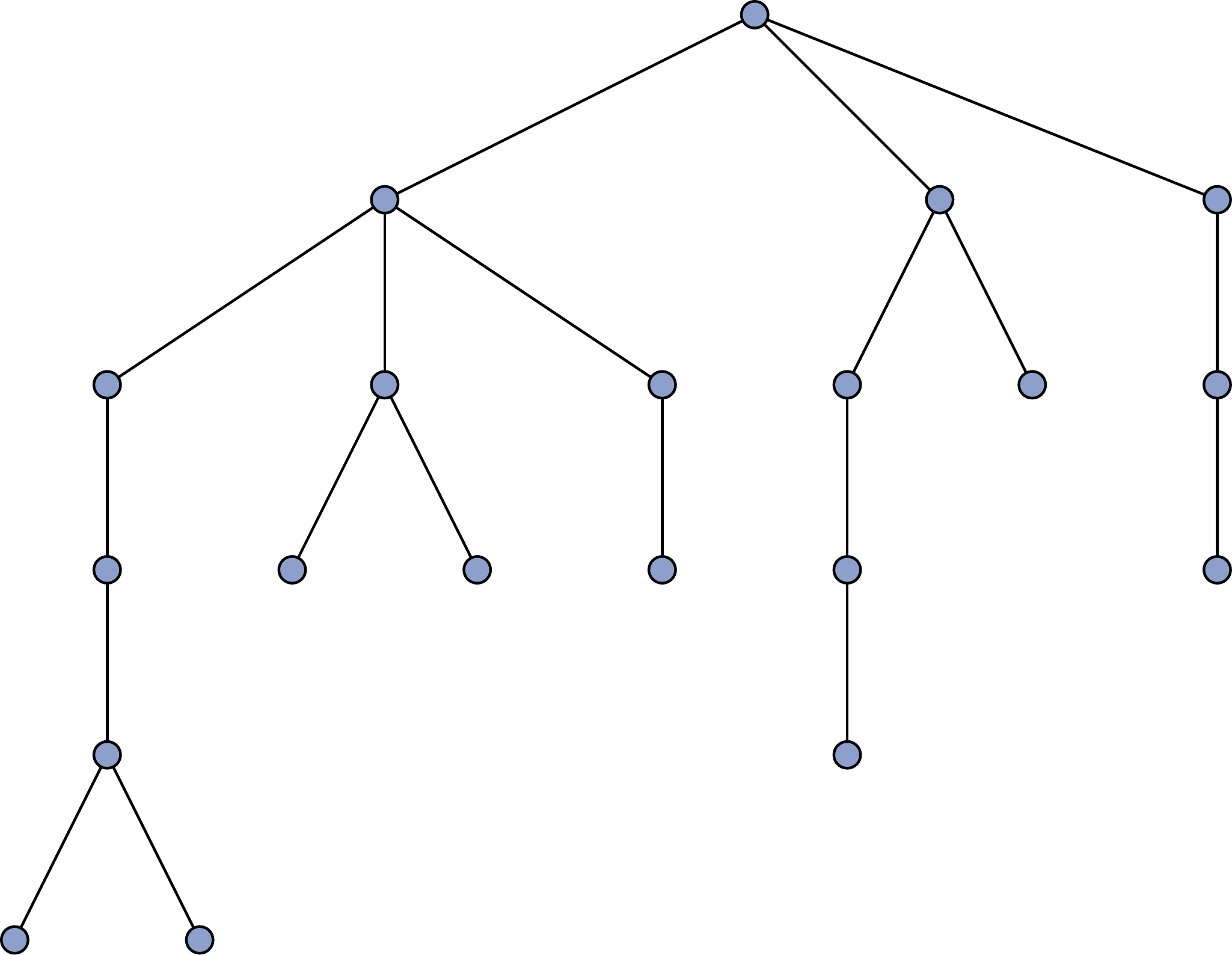}
    \caption{\texttt{Tidier} Layout}
    \label{fig:treeExTidier}
  \end{subfigure}
  \hfill
  \begin{subfigure}[b]{.27\textwidth}
    \centering
    \includegraphics[scale=.2]{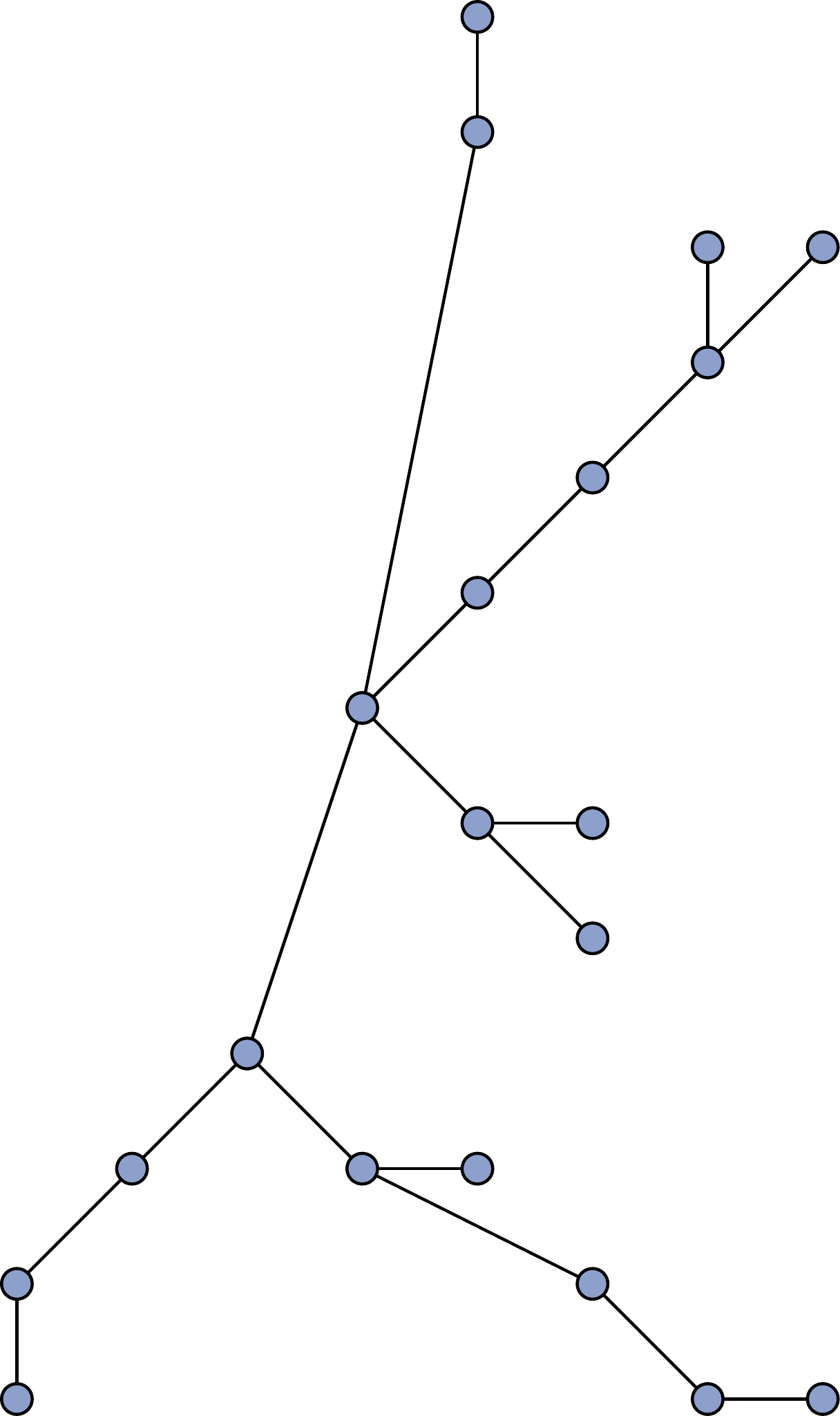}
    \caption{\texttt{Quad} Layout}
    \label{fig:treeExQuad}
  \end{subfigure}
  \hfill
  \begin{subfigure}[b]{.3\textwidth}
    \centering
    \includegraphics[scale=.2]{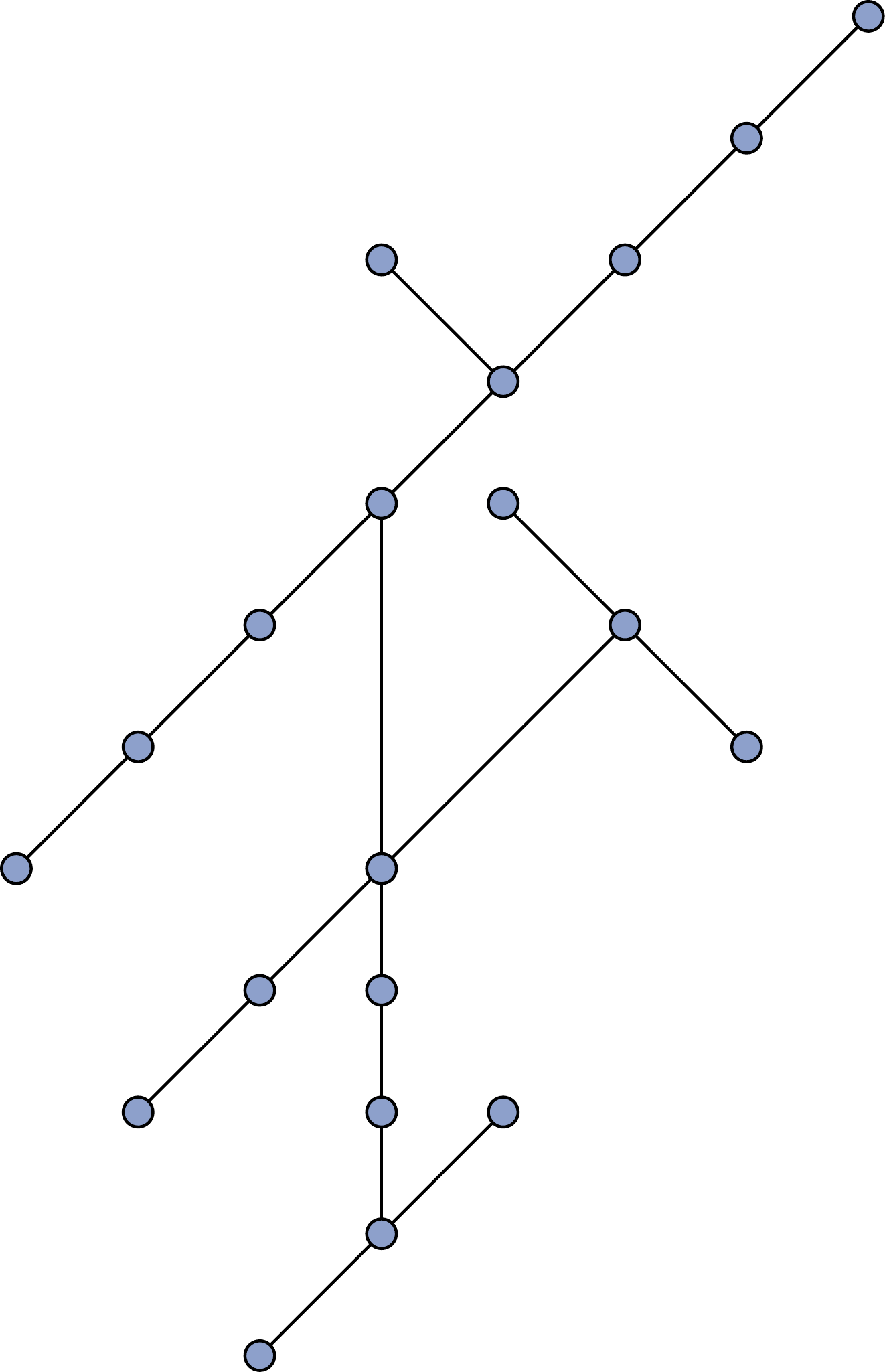}
    \caption{\texttt{FewSegments} Layout}
    \label{fig:treeExFewSeg}
  \end{subfigure}
  \caption{A drawing of a tree with each of the three considered layouts.}
  \label{fig:treeEx}
\end{figure}

\paragraph{Trees.}
For trees, we used three algorithms that produce grid drawings
as illustrated in Fig.~\ref{fig:treeEx}: \texttt{Tidier}, \texttt{Quad}, and \texttt{FewSegments}.
All three algorithms take as input a rooted tree.
Many more examples can be found online~\cite{studyresults}.

The algorithm \texttt{Tidier} was presented by Walker~II~\cite{w-npagt-SPE90} and builds
upon the classic algorithm by Reingold and Tilford~\cite{rs-tdt-TSE81}.
This algorithm produces a drawing on a $O(n)\times O(n)$ size grid that
satisfies three criteria:
\begin{enumerate*}
	\item[(1)] nodes at the same level of the tree should lie along a straight line, and the
straight lines defining the levels should be parallel;
  \item[(2)] a parent should be centered over its offspring;
  \item[(3)] a subtree should be drawn the same way regardless of where it occurs in the tree.
\end{enumerate*}

The algorithm \texttt{Quad} was presented by Rusu et al.~\cite{ryc-psgda-IV08}.
This algorithm allows the user to specify an angular coefficient and draws
edges such that the angles are above the angular coefficient if possible and
evenly spread out otherwise. It also allows
the user to specify how many quadrants may be used to place the children of
a vertex. We chose an angular coefficient of~$22.5^\circ$ and allowed the
algorithm to use all four quadrants.
Higher values for the angular coefficient would lead to poorer angular resolution
in the subtrees; our chosen value gave a well balanced layout for the tree
complexity used in this study. For this algorithm, no bound on the grid size
was given by Rusu et al.

\begin{figure}[t]
  \centering
  \includegraphics{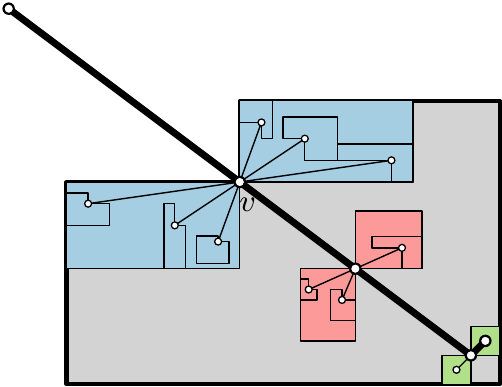}
  \caption{One step of the algorithm by H\"ultenschmidt et al. \cite{hkms-dttfg-wg17}.}
  \label{fig:hueltenschmidt}
\end{figure}

Finally, the algorithm \texttt{FewSegments} is based on the algorithm
by H{\"u}lten-schmidt et al.~\cite{hkms-dttfg-wg17} that draws trees on a
quasi-polynomial grid with a minimum number of segments. On a high level,
that algorithm uses a heavy path decomposition of a tree, which decomposes the
tree in heavy edges and light edges. The paths formed by the
heavy edges are drawn as a single segment. It recursively (guided by the heavy-path decomposition)
embeds each subtree such that the heavy
path of its root is drawn with a vector specified by the parent edge of its
root and all subtrees lie in disjoint regions. The children around a vertex~$v$
that are not connected by a heavy path edge are evenly placed to the top-right of~$v$ with
decreasing $y$-coordinates from left to right and to the bottom-left of~$v$
with increasing $y$-coordinates from right to left with common slopes;
see Fig.~\ref{fig:hueltenschmidt} for an illustration.
We use three heuristics to reduce
the size of the drawing.

The first heuristic is applied during the layout of the tree. When the
algorithm assigns a vector to a subtree, we allow it to increase the length of
the vector slightly such that the new vector is an integer multiple of a smaller
primitive vector. For example, if the algorithm would assign a vector~$(6,11)$,
then this heuristic would change the vector to~$(6,12)$. This implies that the
segments on the heavy path in this subtree do not have to use vectors that
are integer multiples of~$(6,11)$, but only integer multiples of~$(1,2)$.
Although this makes one segment a bit longer, the subtree might use less
area by this change.

In particular, our algorithm takes as an additional parameter some constant~$s\ge 0$.
We place the subtrees in pairs around a vertex from inside to outside
(note that H\"ultenschmidt et al.~\cite{hkms-dttfg-wg17} placed them from outside
to inside, but the order does not matter for their algorithm).
Let~$S$ and~$S'$ be the subtrees to be placed next around~$p$ with the same slope,
let~$(x,y)$ be the vector assigned to the heavy path of~$S$ before applying the heuristic.
Hence, the vector assigned to the heavy path of~$S'$ is $(-x,-y)$.
Note that, if~$p$ has an odd number of light children, then~$S'$ might not exist.
Assume without loss of generality that $x,y>0$.
For all integers $0\le i\le s\cdot x$ and $0\le j\le s\cdot y$
with $(y+i)/(x+i)\le y/x$, the algorithm
computes the width $w_{i,j}$ and the height $h_{i,j}$ for the drawing
of~$S$ if the vector assigned to the heavy path of~$S$ is $\left(x+i,y+j\right)$,
and the width $w'_{i,j}$ and the height $h'_{i,j}$ for the drawing of~$S'$ analogously.
For every choice of~$i$ and~$j$, the algorithm assigns a cost function
\[c(i,j)=i+j+\max\{w_{i,j}+h_{i,j},w'_{i,j}+h'_{i,j}\}.\]
Then, the algorithm chooses~$i^*$ and~$j^*$ such that
\[c(i^*,j^*)=\max_{0\le i\le sx,~0\le j\le sy}c(i,j)\]
and assigns the vector $\left(x+i^*,y+j^*\right)$ to the
heavy path of~$S$ and the vector $\left(x-i^*,y-j^*\right)$ to the heavy path of~$S'$.
By only allowing slopes that are not larger than $y/x$,
we make sure that the edge from~$p$ to the root of~$S$ does not intersect any
already drawn edge; since we are placing the subtrees from inside to outside,
all previously drawn ancestors of~$p$ that are placed to the top-right of~$p$
lie above~$e$ if drawn with vector~$(x,y)$, so drawing~$e$ with a vector of smaller slope
cannot create any crossings. The symmetric argument applies for the edge from~$p$
to the root of~$S'$.
We chose $s=2$ for all tree drawings used in this study.

The second and the third heuristic are applied in alternating order after
a layout has been found. We apply both of them five times.

The second heuristic tries to \emph{compress}
vectors: given a edge~$e$ that is drawn as a vector~$\vec v$ that is an integer
multiple of a primitive vector~$\vec u$, it redraws the tree such that~$e$ is
drawn with the smallest integer multiple of~$\vec u$ without destroying planarity.
This heuristic is applied to every edge in a post-order traversal of the tree.

The third heuristic takes an edge~$e$
that is drawn with a long vector and tries to find a smaller
vector to draw~$e$ and all edges drawn with the same segment as~$e$ such that
the resulting drawing is still planar. This is a more drastic approach and
can change the way a subtree is drawn completely.
Let~$w$ be the width of the current drawing and let~$h$ be its height.
Let~$S$ be a subtree with root~$r$ and parent~$p$ such that $(p,r)$ is not a
heavy edge, it is drawn with vector $(x,y)$, and $|x|+|y|>(w+h)/5$.
Depending on the number of children of~$r$, there might be another subtree~$S'$
with parent~$p'$ such that the edge~$(p,r')$ drawn with vector $(-kx,-ky)$
for some integer~$k$.
For all integers $i,j$ with $0<|i|+|j|<|x|+|y|$ (in order of rising $|i|+|j|$),
the algorithm
computes a drawing of the tree where $(p,r)$ is drawn with vector $(i,j)$,
the vector $(i,j)$ is assigned to the heavy path of~$S$,
the edge $(p,r')$ is drawn with vector $(-i,-j)$, the vector $(-i,-j)$ is
assigned to the heavy path of~$S'$, and the drawing of all other edges does not
change. If the resulting drawing is planar and its width and height are not
higher than those of the original drawing, then the algorithm keeps the drawing;
otherwise, it uses the next values of~$i$ and~$j$ until all of them are used.
This heuristic is again applied to every edge in a post-order traversal of the tree,
but only those that fulfill the required conditions.

\begin{figure}[t]
  \centering
  \begin{subfigure}{.45\textwidth}
    \centering
    \includegraphics[scale=.25]{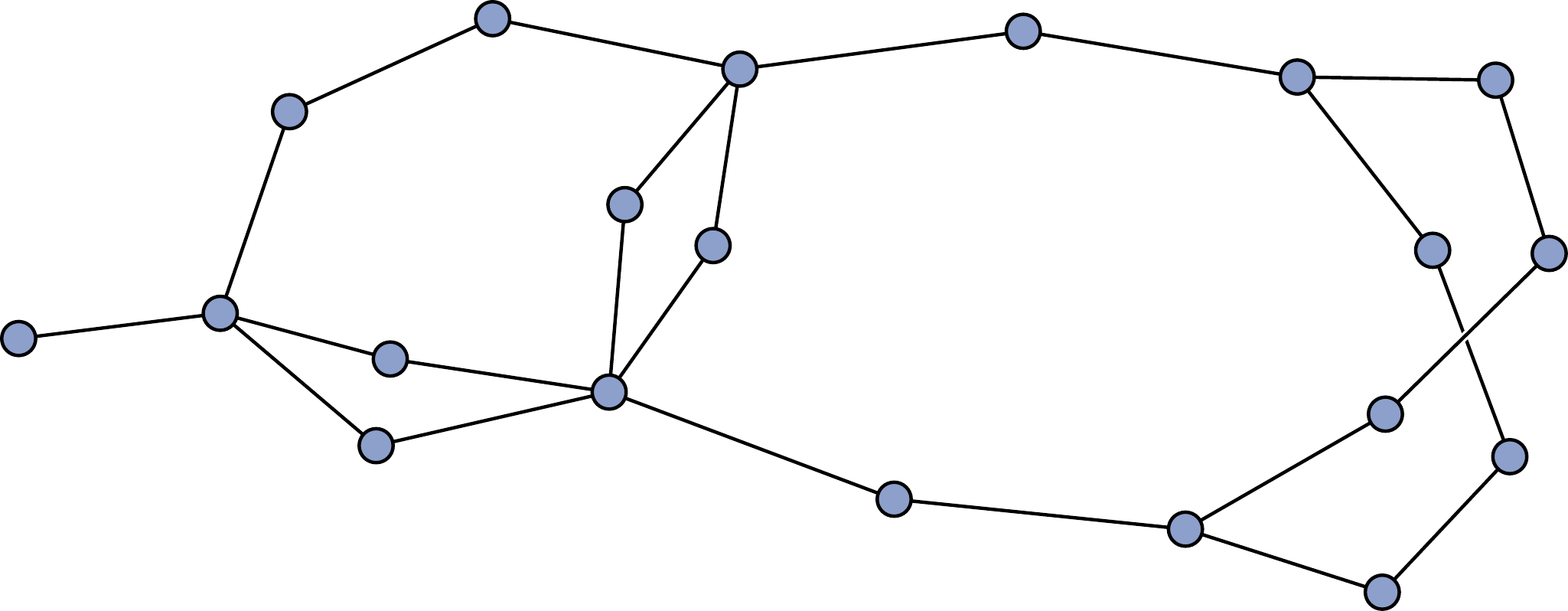}
    \caption{\texttt{ForceDir} Layout}
    \label{fig:forceExForceDir}
  \end{subfigure}
  \hfill
  \begin{subfigure}{.45\textwidth}
    \centering
    \includegraphics[scale=.25]{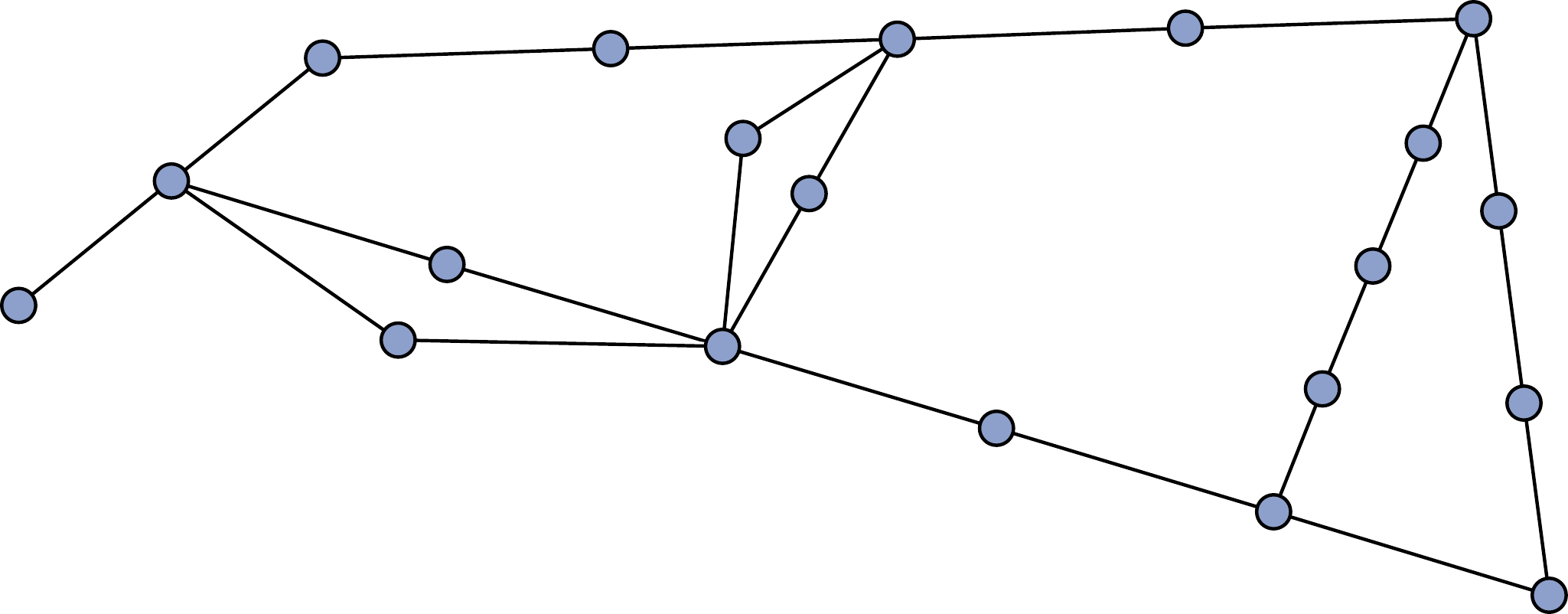}
    \caption{\texttt{FDFewSeg} Layout}
    \label{fig:forceExFewSeg}
  \end{subfigure}
  \caption{ROME graph \emph{2282.20} using both force-directed layouts.}
  \label{fig:forceEx}
\end{figure}

\paragraph{Graphs.}
For sparse graphs we used the algorithms \texttt{ForceDir} and
\texttt{FDFewSeg}; example drawings are provided in Fig.~\ref{fig:forceEx} and
online~\cite{studyresults}. The former is an implementation of the
spring embedder by Fruchterman and Reingold~\cite{fr-gdfdp-SPE91}.
This algorithm computes a force between each pair of vertices.
If there is an edge between two vertices, then there is an \emph{attractive force}
$f_a(d)=d^2/k$ between them, where~$d$ is the distance between the vertices and~$k$ is their
optimal distance defined as~$k=C\sqrt{A/n}$, where~$C$ is some constant,~$A$
is the maximum area of the drawing, and~$n$ is the number of vertices in the
graph. If there is no edge between two vertices, then there is a \emph{repulsive
force}~$f_r(d)=-k^2/d$ between them. By an addition of these forces at every
vertex~$v$, we obtain a \emph{movement}~$\Delta v$ of the vertex described
by a 2-dimensional vector. Fruchterman and Reingold use simulated
annealing to control the movement of the vertices such that the adjustments
become smaller over time and the algorithm terminates.

The \texttt{FDFewSeg} algorithm is an extension of the \texttt{ForceDir} spring
embedder that we developed for this paper.
It takes as an additional input a set of edge-disjoint
paths~$\mathcal P$. First, the movement~$\Delta v$
for every vertex is computed. Let~$v_0,\ldots,v_k=P\in\mathcal P$.
The algorithm places the internal vertices~$v_1,\ldots,v_{k-1}$ evenly spaced onto the
segment between the endvertices~$v_0$ and~$v_k$. To this end, for every vertex~$v_i,0<i<k$,
the movement becomes
\[\Delta v_i=\frac{n-i}{n}\left(v_0+\Delta v_0\right)+\frac{i}{n}\left(v_k+\Delta v_k\right) - v_i.\]
Note that this procedure does not necessarily draw all paths in~$\mathcal P$
as segments: if a vertex~$u$ of a path is an internal vertex of another path
that is processed later, then~$u$ will be moved away from its path segment.
Hence, the user should input paths in an order that avoids this problem:
Let~$\mathcal P=P_1,\ldots, P_m$. Every vertex can be the internal
vertex of at most one path. If some vertex If~$v$ is an internal vertex of a
path~$P_i$ and the endvertex of another path~$P_j$,
then~$P_j$ has to be input after~$P_i$.

For the creation of the drawings for the user study, we picked the paths
to be placed on segments manually; most of the time, we picked paths that are
(somewhat) close to being segments in the drawing of the \texttt{ForceDir} algorithm.

We chose not to use an automated way to select the paths for the \texttt{FDFewSeg}
algorithm. The aim of this paper is to compare the aesthetic criteria of the drawing styles,
which could be negatively influenced by a ``bad'' path selection algorithm.
Automated strategies for selecting paths only become relevant if the
aesthetic criterion is worthwhile. We note that, since there is some hint
on the criterion being worthwhile, finding a good automated strategy is something
to be done in future work.

\section{Hypotheses}

We designed a user study to compare aesthetics and legibility of drawings produced by the above-described algorithms.
For the study we posed the following four hypotheses:

\begin{enumerate}[label=\bf H\arabic*.,nolistsep]
	\item  For trees, the aesthetics ranking is
    \begin{itemize}[nolistsep,leftmargin=0mm]
	\item[\bf a] \texttt{Tidier} $>$ \texttt{FewSegments}
    $>$ \texttt{Quad} for people with a mathematics or computer science background, and
    \item[\bf b] \texttt{FewSegments} $>$ \texttt{Tidier} $>$ \texttt{Quad} for people
    from a different background.
    \end{itemize}
  \item For trees, path finding is easiest with the \texttt{FewSegments}
    layout, followed by \texttt{Tidier}, and hardest with \texttt{Quad}.
	\item For sparse graphs, the \texttt{ForceDir} layout is more aesthetically pleasing
    than the \texttt{FDFewSeg} layout.
	\item For sparse graphs, path finding is easier
    with the \texttt{FDFewSeg} layout than with the \texttt{ForceDir} layout.
\end{enumerate}
For Hypothesis~H1, we expect that the uniformity of the \texttt{Tidier}
and the \texttt{FewSegments} layout make them preferred over the \texttt{Quad}
layout. For mathematicians or computer scientists, we expect that the
\texttt{Tidier} layout is preferred, since it creates a drawing in the standard way
that trees are drawn in the literature. For people with different background,
we expect that the \texttt{FewSegments} layout is preferred, because it seems to be
more schematic. Hence, this hypothesis is split into two parts, H1a and H1b.

The same idea underlies Hypotheses~H2 and~H4: placing paths onto
few segments makes it easier for the user to follow a path between two nodes,
since the eye only has to move along few directions and can traverse several
nodes quickly along a segment. Evenly distribution the nodes along a path
in the force-directed layout should help the reader to quickly determine the
number of nodes on a segment and thus to judge the combinatorial length of such
a path.

For Hypothesis~H3, we think that the smooth curves in the \texttt{ForceDir}
layout look nicer to a reader than the drawings of the \texttt{FDFewSeg}
layouts because the latter ones can have sharp corners at the meeting
point of two path segments; for example, Bar and Neta \cite{BarNeta} argue that sharp corners have a negative effect on aesthetics as such bends are identified with threat.
On the other hand, Vessel and Rubin \cite{VesselRubin} studied the objectiveness of taste---their conclusion is that there is typically agreement for natural images, abstract depictions are influenced more by individual taste.
Though they cannot fully be eliminated, we believe that the uniformity of graphical presentation may mitigate personal preferences to allow for investigating an overall agreement in aesthetics.

\section{Experimental design}

\paragraph{Selecting tasks.}
We used two tasks in the study: \textbf{Aesthetics} and \textbf{Query}. We created different
graphs for each task. For the Aesthetics task, we showed the participant one drawing
for each layout of the same graph next to each other. The order of the
drawings was determined randomly. The participant was asked to determine
a ranking on the aesthetics of the drawings by clicking on them in the desired
order.

We used different Query tasks based on the graph class.
We showed the participant one drawing at a time.  Over the course of the experiment, every graph was presented to
the participant once with each layout.

For the sparse graphs, we asked the participants to find the shortest path
between two randomly marked vertices that have distance at least~3 (the pair of vertices was the same
for each layout and each participant). The participant solved this task by clicking on the vertices (or edges) in the order
that they appear on this path. To make sure that a participant does not get stuck on
a question, we allowed them to submit their answer even if no path was found.
We helped the participant with this task by marking (in a different color) the valid nodes and edges they
can click on, which are those that are adjacent to the endpoint of one of
the two paths starting in the two marked vertices.

For trees, shortest paths are uniquely defined which makes it unsuitable
as a task. Hence, we asked the participant to find the furthest pair of vertices, that is,
the pair of vertices such that the distance between them is maximized.
Like finding shortest paths in general graphs, this also requires the participant to inspect several paths to determine the answer.
The participant then had to click on the vertices that they determined as the furthest
pair.

\paragraph{Generating stimuli.}
All stimuli and their drawings have been made available online~\cite{studyresults}.
For trees, we have the following two variables for the stimuli:
\begin{itemize}[nolistsep]
	\item \textbf{Size.} Two different sizes: (1) 20 nodes and (2) 40 nodes.
  \item \textbf{Depth.} Three different tree depths as defined by the length
    of the longest root--leaf paths: (D) \emph{deep} trees of depth~8
    for size~1 and of depth~14 for size~2, (B) \emph{balanced} trees of
    depth~5 for size~1 and of depth~9 for size~2, and (W)~\emph{wide}
    trees of depth~3 for size~1 and of depth~5 for size~2.
\end{itemize}
We use rejection sampling to construct random trees of given size and depth.
We first create a uniformly distributed random Pr{\"u}fer
sequence~\cite{p-nbsp-AMP18} and the corresponding labeled unrooted tree.
We always choose the vertex with label~$1$ as the root to create a
rooted tree and then check whether it
has the given depth. It is known that Pr{\"u}fer sequences provide a bijection
between the set of labeled trees on~$n$ vertices and the set of sequences
of~$n-2$ integers between~$1$ and~$n$. Hence, this algorithm gives us
uniformly distributed random trees of a given depth. For each size and depth,
we created four different graphs for the Aesthetics task and two different
graphs for the Query task. This gives us $2\cdot 3\cdot 4=24$ graphs for the
Aesthetic tasks (4 repetitions) and $2\cdot 3\cdot 2=12$ graphs for the Query task (2 repetitions).

For the sparse graphs, we have the following two variables for the stimuli:
\begin{itemize}[nolistsep]
	\item \textbf{Size}. Two different sizes: (1) 20 nodes and (2) 40 nodes.
  \item \textbf{Type}. Two different types: (A) graphs from the ROME library
    and (B) random graphs.
\end{itemize}
For graphs of type~A, we randomly picked graphs of the given size from the
ROME library~\cite{wdglttv-ecfgd-CG97} that consists of 11,535 sparse, but not
necessarily planar, graphs with~10 to~100 vertices. For graphs of type~B,
we created a random graph by creating a number of nodes specified by the size
and picking 30 random edges for graphs of size~1 and~60 random edges for graphs
of size~2; we used the resulting graph if and only if it is connected.
Our choice leads to comparatively sparse but not necessarily planar graphs, to ensure legible layouts for the user study.
For each size and type,
we again created four different graphs for the Aesthetics task and two different
graphs for the Query task. This gives us $2\cdot 2\cdot 4=16$ graphs for the
Aesthetic tasks (4 repetitions) and $2\cdot 2\cdot 2=8$ graphs for the Query task (2 repetitions).

We stored the graphs as JSON files, which contained the coordinates of the
vertices and the set of edges. During the study, the graphs were drawn
using the JavaScript library \emph{D3.js}~\cite{boh-d3-TVCG11} as SVG figures to allow arbitrary resizing.
The nodes were drawn using blue circles. Links were drawn in black with a small
halo to increase separability between crossing links. The selected vertices
and links in both Query tasks were marked in green and the selectable
vertices and links in the shortest path task were marked in light blue.

\paragraph{Further considerations.}
For trees, we created~$24$ stimuli  for the Aesthetics task and~$36$ stimuli
for the Query task (one per graph and layout algorithm). For the sparse graphs, we
created~$16$ stimuli for the Aesthetics task and~$16$ for the Query task.
This gives us~$92$ stimuli in total. This is beyond what is reasonable for
an online study, assuming 15 to 25 seconds per trial. Since the study has
two different graph classes with different tasks, we used the graph class
as a between-subjects measure. This still leaves~$60$ stimuli for the tree
tasks. Since the size of a graph is very likely to be an overall factor
by the larger difficulty of the Query task on a larger graph, we used the
size as an additional between-subjects measure for trees. This way, we obtain
three groups of stimuli: (1) $30$ stimuli for trees of size~$1$, (2)~$30$
stimuli for trees of size~$2$, and (3)~$32$ stimuli for sparse graphs.
A pilot study showed a completion time of about~$15$ minutes for each
group.

We first show the Aesthetics task and then
the Query task. We did this such that the participant does not get a bias for a
specific drawing style based on the difficulty of the Query task and instead of
the most aesthetic one picks the one that they preferred in the Query section.
Though explicitly asking for visual preference could bias performance in the following Query section,
we expect this effect to be negligible, since only one drawing is shown at any given time; and in any case less strong than the potential bias if the sections were to be inverted.

In order to account for learning effects, the orders of the stimuli for each task were
randomized for each participant. Before each stimulus, the participant was given a
pause screen to reduce memory effects and at the same time allow them to pace
themselves and reduce the possible impact of interruptions.
The participants received one example question with an answer revealed after providing one, from which they could go back to task description, to ensure that the task was understood before starting the actual questions.
We opted not to provide a longer series of training questions to keep time investment to a minimum.

\paragraph{Setup.}
We developed our user study with PHP and the JavaScript library \emph{D3.js}.
The study was hosted on a web server\footnote{http://tutte.fernuni-hagen.de/web/userstudy/fewarcs}
and the data was stored in a MySQL database. Since the questionnaire was conducted
online, we had no control over many parts of the experimental environment, e.g.,
device, pointing device, operating system, browser, screen size, interruptions.
We asked the participants to fill in the questionnaire using a desktop or laptop
computer, not a tablet or phone, and to use the pointing device they are
most comfortable with. To make sure that the browser is suitable to run the
questionnaire, the participants first had to set a slider to the value depicted in
an SVG figure.
We could not control the screen size, resolution, or distance of the participant
to their screen, so we let the participant control the scale of the web page by
providing a \emph{Shrink} and a \emph{Grow} button. Further,
we asked them to put their browser in full-screen mode
to reduce distractions. We requested the participants to not engage in other
activities during the questionnaire and to minimize interruptions. At the end of the study, we asked participants to specify if any interruptions occurred during the questionnaire.

We recruited the voluntary participants of the user study using a mix of
mailing lists, social networks, and social media.
Some background and preference information was asked upon completion, although this remained optional for what may considered sensitive
information (age, gender, country of residence).

\section{Results}
The data set for the analysis as well as all stimuli have been made available
online~\cite{studyresults}.
In total, 84 people volunteered and completed the online questionnaire,
which was open for participation for two weeks.
We inspected all comments left by participants.
One participant had a longer break during one of the questions, rendering this particular question unsuitable for the analysis. As to maintain a balanced design to allow for stronger analysis methods, we
excluded this participant from the analysis. This gave us~21 participants for both group~1
and group~2, and~41 participants for group~3. Of the~83 participants,~75 provided
their age with an average of~$36.87$, a median of~$33$, and a standard deviation of~$11.41$.
In terms of country of residence, a majority of
the participants live in Europe (63), predominantly in Germany (42).

For Hypothesis~H1, we expected different results for people from mathematics
or computer science background (H1a) than for people from a different background (H1b).
However, we asked for the background of the participants only after they
completed the questionnaire, so we could not influence the distribution
between the three different groups. Unfortunately, this resulted in only
two participants without a mathematics of computer science background to
be assigned to the groups for drawings of trees, so we could not claim a
significant effect for H1b. Hence, we started a follow-up study that consisted
only of the aesthetics task on tree drawings, but for trees of both sizes.
We advertised this study on Reddit\footnote{https://www.reddit.com/r/SampleSize/}
to specifically get people from more diverse backgrounds. This resulted in~24
additional participants, 14 of them did not have a mathematics or computer science
background. Of these participants,~23 provided their age with an average of~$27.65$,
a median of~$25$, and a standard deviation of~$6.63$.
The countries of residence also turned out to be more diverse, with 8 participants
from Europe, 7 from USA, 4 from Canada, 2 from Australia, 1 from Japan,
and 2 unspecified. The results from the follow-up study helped us
to evaluate Hypothesis~H1 more precisely, compared to the conference version where that data
was not available yet.

\begin{figure}[t]
  \centering
  \includegraphics[page=1,scale=.75]{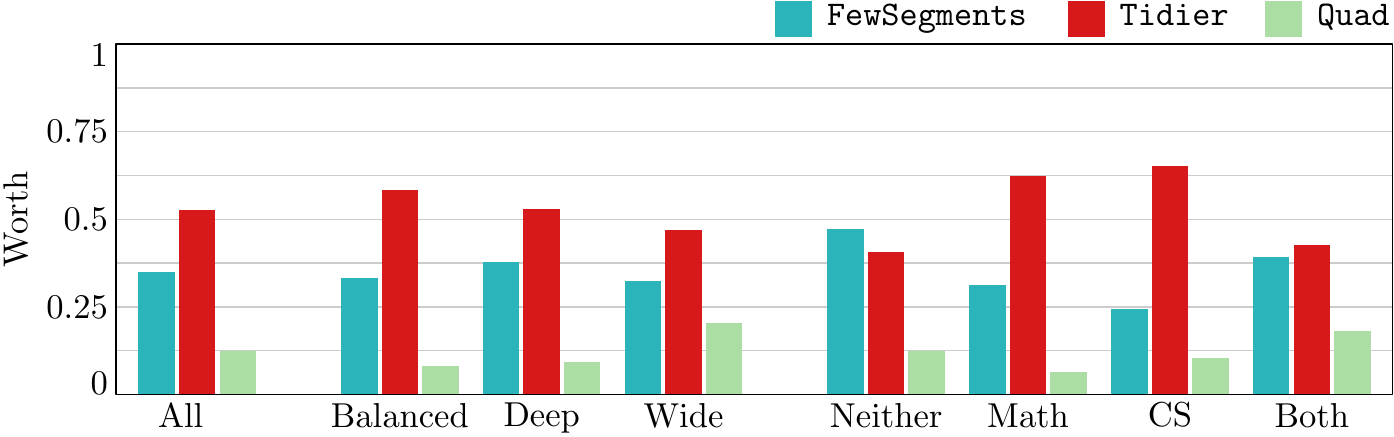}
  \caption{Worth scores of the three tree layout methods: overall and partitioned by tree balance or participant background.}
  \label{fig:h1-worthscores}
\end{figure}

\paragraph{Hypothesis~H1.}
For the tree aesthetics task, we had~42 participants from groups~1 and~2 and
each of them was shown~12 stimuli, and we had~24 participants from the follow-up
study and each of them was shown~24 stimuli. This gave us a total of~1,080 rankings
between the three layouts.
We used loglinear Bradley-Terry (LLBT) modeling~\cite{hd-prefm-JSS11} of the
3,240 pairwise aesthetic preference comparisons to produce ranked \emph{worth}
scores for each of the three layouts. The worth score allows the
consistency of preference to be assessed in forming an overall
ranking of the three classes. Fig.~\ref{fig:h1-worthscores} shows the ranking of
the three layouts in terms of aesthetic preference, broken down by the
balance of the graph and by the background of the participants.

Consistently, the \texttt{Quad} layout was considered as the least aesthetic
tree layout. Over all answers, the \texttt{Tidier} layout performed the best.
There was some effect based on the balance of the graph. For each balance,
we received~360 rankings. However, the ranking for each balance is the same
as the overall ranking, they differ only in the effect size.
For \emph{balanced} trees, the effect is the largest with worth
scores~$0.5845$ (\texttt{Tidier}),~$0.3337$ (\texttt{FewSegments}),
and~$0.0817$ (\texttt{Quad}). For \emph{deep} trees, we have worth
scores~$0.5288$ (\texttt{Tidier}),~$0.3782$ (\texttt{FewSegments}),
and~$0.0931$ (\texttt{Quad}). For \emph{wide} trees, the effect is the smallest,
but still significant, with worth
scores~$0.4709$ (\texttt{Tidier}),~$0.3238$ (\texttt{FewSegments}),
and~$0.2053$ (\texttt{Quad}).

The hypothesis was split into two parts, depending on the background of the
participants. Let us first consider the participants with a mathematics or computer science background.
There were 286 rankings by 19 people with a mathematics background,
but not computer science; for those, the \texttt{Tidier} layout
was clearly preferred with a worth score of~$0.6231$ over~$0.3136$.
There were~648 rankings by~45 participants with a computer science background,
but not mathematics; for those, the \texttt{Tidier} layout
was also clearly preferred with a worth score of~$0.6533$ over~$0.2434$.
There were~192 rankings by~13 participants with both mathematics and computer science
background; those slightly preferred the \texttt{Tidier} layout
with a worth score of~$0.4273$ over~$0.3914$.
Overall, this suggests that the suspected preference of the \texttt{Tidier} layout over the \texttt{FewSegment} layout, and of both layouts over the \texttt{Quad} layout exists;
hence, we accept Hypothesis~H1a.

Hypothesis~H1b is about participants from neither
mathematics nor computer science background. There were~336 rankings by~15
people with neither a mathematics background, nor a computer science background.
These participants indeed preferred the \texttt{FewSegment} layout over the \texttt{Tidier} layout
with a worth score of~$0.4714$ over~$0.4054$. This therefore also confirms Hypothesis~H1b, so we may accept Hypothesis H1.

\begin{figure}[t]
  \centering
  \includegraphics[page=1,scale=.75]{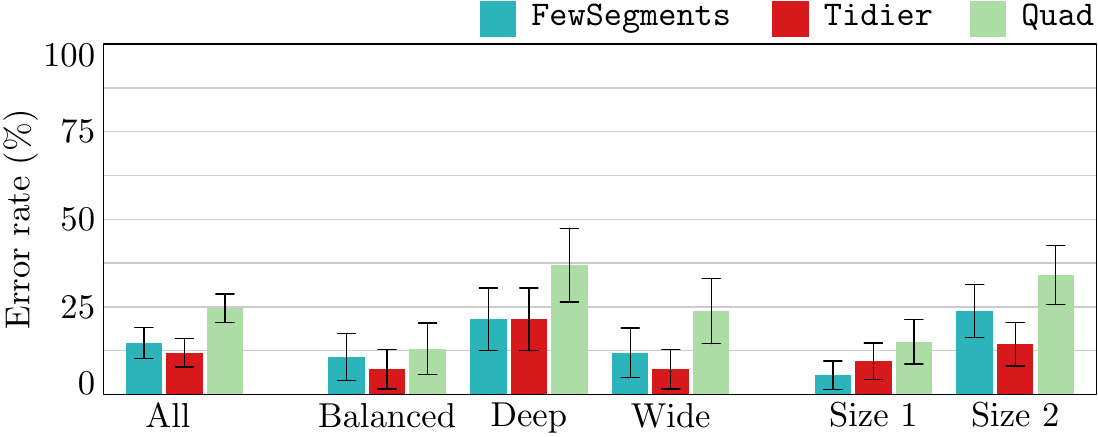}
  \smallskip

  \includegraphics[page=1,scale=.75]{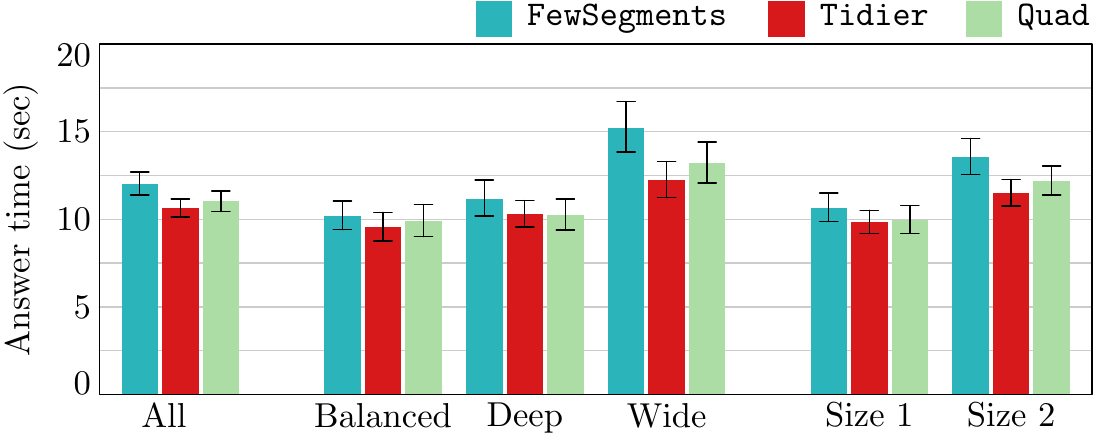}
  \caption{Error rates and answer times for finding a furthest pair for the three tree layout methods: overall and partitioned by tree balance or size group. Error bars indicate 95\% confidence intervals.}
  \label{fig:h2-errorrates}
\end{figure}

\paragraph{Hypothesis~H2.}
For the tree query task, we had~42 participants from groups~1 and~2 and
each of them was shown~18 stimuli. This gave us a total of~756 tasks
between the three layouts with~252 tasks per layout.
We analyzed the error rates for finding a furthest pair for the three tree
layouts defined by the difference of the distance between the picked pair and
the distance between a furthest pair in the graph,
broken down by the balance and by the size of the trees.
The maximum response time was~53 seconds, so we did not have to exclude
any participants.
Fig.~\ref{fig:h2-errorrates} shows the error rates and
the answer times.

We used a two-way RM-ANOVA test to analyze the effects of the layouts, tree balance,
tree size, and their interaction. We used the logarithm of the response
times to normalize the distribution. For the error rate, there are no
interaction effects between layout, balance, and size. The analysis showed
a weak effect of the layout on the error rate ($F(2,80)=3.636$, $p<0.05$).
A post-hoc Tukey HSD test with Bonferroni adjustment showed a significant
difference between layout \texttt{Quad} and \texttt{FewSegments} in favor of
\texttt{FewSegments} ($p<0.01$) and a significant
difference between layout \texttt{Quad} and \texttt{Tidier} in favor of
\texttt{Tidier} ($p<0.001$), but no significant difference between layout
\texttt{Tidier} and \texttt{FewSegments}. Further, a post-hoc test showed
a weak difference between the tree sizes ($p<0.05$) in favor of
smaller trees.
For small trees, there is some evidence that \texttt{FewSegments} outperforms \texttt{Tidier} ($p < 0.05$); for large trees the error rate seems lower for \texttt{Tidier}, though no statistically significant effect was found ($p > 0.15$).
We conclude that the layouts \texttt{FewSegments} and \texttt{Tidier}
perform better than the layout \texttt{Quad}, while the participants performed
better on small trees than on large trees.

For the response time, there is some interaction between
tree size and tree balance ($F(4,160)=2.524$, $p<0.05$), so we split
according to sizegroup for further analysis.
For small trees, there are no interaction effects between layout and tree balance.
The analysis showed a very weak effect of layout ($F(2,40)=2.523$, $p<0.1$) on response time.
A post-hoc test showed a very weak difference between layout \texttt{Tidier}
and \texttt{FewSegments} in favor of \texttt{Tidier} ($p<0.1$) and no
significant difference between the other two layout pairs.
We conclude that the participants performed slightly faster for the
\texttt{Tidier} layout than for the \texttt{FewSegments} layout

For large trees, there are also no interaction effects between layout and tree balance.
The analysis showed significant effect of layout ($F(2,40)=9.667$, $p<0.001$) on response time.
A post-hoc test showed a weak difference between layout \texttt{Quad}
and \texttt{FewSegments} in favor of \texttt{Quad} ($p<0.05$) and a
significant difference between layout \texttt{Tidier}
and \texttt{FewSegments} in favor of \texttt{Tidier} ($p<0.001$).
We conclude that the participants performed slightly faster for the
\texttt{Quad} layout than for the \texttt{FewSegments} layout
and significantly faster for the
\texttt{Tidier} layout than for the \texttt{FewSegments} layout.

Since the error rate was smaller for the \texttt{FewSegments} and \texttt{Tidier}
layouts than for the \texttt{Quad} layout, but the response time for \texttt{FewSegments}
was worse than for the other two, we can only partially accept Hypothesis~H2:
the layouts \texttt{FewSegments} and \texttt{Tidier} both outperform the
layout \texttt{Quad}, but the layout \texttt{Tidier} outperforms the layout
\texttt{FewSegments}.

Though not initially hypothesized, we also found a significant effect of the
tree balance on the error
rate ($F(2,80)=14.867$, $p<0.001$).
A post-hoc test showed a significant
difference between tree balances \emph{balanced} and \emph{deep} in favor of
\emph{balanced} ($p<0.001$) and a significant
difference between tree balances \emph{wide} and \emph{deep} in favor of
\emph{wide} ($p<0.001$), but no significant difference between tree
balances \emph{wide} and \emph{balanced}.

For small trees, we found a significant effect of balance ($F(4,80)=65.07$, $p<0.001$)
on the response time. There is a significant difference between both balances \emph{wide} and
\emph{balanced} and the balance \emph{deep} in favor of the former ($p<0.001$ for
both).

For large trees, the analysis showed a
significant effect of balance ($F(4,80)=1.808$, $p<0.001$) on the response time.
A post-hoc test showed a significant difference between balance \emph{wide} and
\emph{balanced} in favor of \emph{balanced} ($p<0.001$),
a significant difference between balance \emph{wide} and
\emph{deep} in favor of \emph{deep} ($p<0.01$), and
a weak difference between balance \emph{deep} and
\emph{balanced} in favor of the \emph{balanced} ($p<0.05$).

This analysis suggests that it is easier to find a furthest pair in \emph{balanced}
and \emph{wide} trees. We believe that this effect comes from a correlation
between the distance of a furthest pair and the depth of the tree
and that finding a furthest pair seems to be easier the shorter the distance
between them is. However, since we have no hypothesis on this behavior,
we cannot claim the statistical effect as a strong evidence of an effect.

\begin{figure}[t]
  \centering
  \includegraphics[page=2,scale=.75]{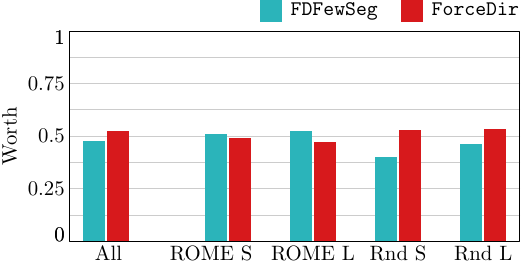}
  \caption{Worth scores of the two graph layout methods: overall and partitioned by type.}
  \label{fig:h3-worthscores}
\end{figure}

\paragraph{Hypothesis~H3.}
For the sparse graph aesthetics task, we had~41 participants from group~3 and
each of them was shown~16 stimuli. This gave us a total of~656 rankings
between the two layouts. We again used LLBT modeling
of the~656 pairwise aesthetic preference comparisons to produce ranked worth
scores for both layouts. Fig.~\ref{fig:h3-worthscores} shows the ranking of
the both layouts in terms of aesthetic preference, broken down by the
graph class and the size of the graph.

Over all~656 rankings, the \texttt{ForceDir} layout
was preferred with a worth score of~$0.5246$ over the \texttt{FDFewSeg}
layout with a worth score of~$0.4754$.
For graphs from the ROME library, the \texttt{FDFewSeg} layout was slightly preferred
by the participants. For small ROME graphs, the layouts were perceived similar, with
worth scores of~$0.5089$ (\texttt{FDFewSeg}) and~$0.4911$ (\texttt{ForceDir}),
but for large ROME graphs, there is a clearer (although still small) preference
for \texttt{FDFewSeg} (worth score~$0.5268$) over \texttt{ForceDir}
(worth score~$0.4732$). On the other hand, for randomly generated graphs, there
is a clear preference for the \texttt{ForceDir} layout over the \texttt{FDFewSeg}
layout. The effect is the largest for small random graphs with a worth score
of~$0.5977$ over~$0.4023$; for large random graphs, the worth scores
are~$0.5357$ (\texttt{ForceDir}) and~$0.4643$ (\texttt{FDFewSeg}).
Hence, we accept Hypothesis~H3, although we remark that for the real-world
graphs from the ROME library the layouts performed similary with a
slight preference for \texttt{FDFewSeg}.

\begin{figure}[b]
  \centering
  \includegraphics[page=2,scale=.68]{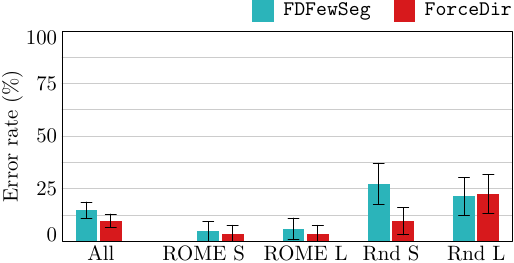}
  \hfill
  \includegraphics[page=2,scale=.68]{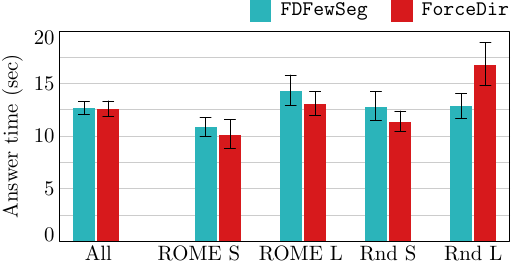}
  \caption{Error rates and response times for finding a shortest path for the two layout methods: overall and partitioned by graph type. Error bars indicate 95\% confidence intervals.}
  \label{fig:h4-errorrates}
\end{figure}

\paragraph{Hypothesis~H4.}
For the sparse graph query task, we had~41 participants from group~3 and
each of them was shown~16 stimuli. This gave us a total of~656 tasks
between the two layouts with~328 tasks per layout.
We analyzed the error rates for finding a shortest path for the two
layouts defined by the difference between the length of the selected
path and the length of a shortest path,
broken down by the four graph types (ROME small, ROME large, Random small, Random large).
Fig.~\ref{fig:h4-errorrates} shows the error rates and
answer times by the participants.

We used the same analysis as for the tree query task.
For the error rate, there is some interaction between layout
and graph type ($F(3,120)=3.313$, $p<0.05$), so we split according to
graph type.
For \emph{Random small} graphs, we found a significant
difference between the layouts in favor of \texttt{ForceDir} ($F(1,40)=9.949$, $p<0.01$);
for the other graph types, there is no significant effect of the layouts.

For the response time, there is a strong interaction between
layout and graph type ($F(3,120)=21.06)$, $p<0.001$), so we split
according to graph type. For \emph{ROME small} ($F(1,40)=9.317$, $p<0.01$)
and \emph{Random small} graphs ($F(1,40)=7.474$, $p<0.01$), there is a significant
effect of the layouts on the response time in favor of \texttt{ForceDir}.
For \emph{ROME large} graphs, there is a very weak
effect of the layouts on the response time in favor of \texttt{ForceDir} ($F(1,40)=3.901$, $p<0.1$).
For \emph{Random large} graphs, there is a significant
effect of the layouts on the response time in favor of \texttt{FDFewSeg} ($F(1,40)=24.56$, $p<0.001$).

Since the \texttt{ForceDir} layout outperformed the \texttt{FDFewSeg}
on three of the four graph layouts, we have to reject Hypothesis~H4 in general.
However, the \texttt{FDFewSeg} performed better on
large random graphs, so there is some evidence that this layout can give
better results if the input graph has many vertices.
The reason for this may be that
the ROME graphs tend to have many degree-2 vertices. Similar effects can  be observed for the
small random graphs. Consequently, paths become easily traceable for these instances
even if drawn with many segments.

\section{Conclusion}

We compared various graph layout algorithms to assess the effect of low visual complexity on aesthetics and performance.
We have confirmed Hypothesis~H1: people with a math or computer science background tend to prefer the classical top-down layout for trees, and
people with no such background prefer the layouts produced by the algorithm assuring low visual complexity.
We have also partially confirmed Hypothesis~H2, by discovering evidence that finding a furthest pair is the easiest with the classical tree layout.
We confirmed Hypothesis~H3: for sparse graph the traditional force-directed layout is
more aesthetically appealing than its modification to reduce the visual complexity, although
there is some evidence that for real-world graphs the effect is very small and
might even be in favor of the layout with small visual complexity.
We rejected Hypothesis~H4 in general, but rather found that it is typically easier to find the
shortest paths between two nodes with the traditional force-directed layout than the modification, though our hypothesis was found to hold for large random graphs. This leaves the possibility open that using few segments
can be beneficial for graphs that are more intertwined.

In short, our findings suggest that visual complexity may positively influence aesthetics, depending on the background of the observer, as long as it does not introduce unnecessarily sharp corners. Hence, drawings trees with few segments give a more schematic alternative over the classic drawing style without the risk of harming the aesthetic perception. However, few-segment drawings tend not to improve task performance. It is worth noting that we did not provide training to our participants, as to suggest how the segments can for example help to easily assess the length of a subpath. Providing such clues may have a positive effect on the performance, but at the same time would also result in an unfair comparison, if no training or strategies for the traditional layout were to be suggested.

\paragraph{Acknowledgments}
The authors would like to thank all anonymous volunteers who participated in the presented user study
and the anonymous reviewers for their valuable input.

\clearpage
\bibliographystyle{abbrvurl}
\bibliography{abbrv,fewarcs}

\end{document}